\documentclass[12pt]{article}
\usepackage{amsmath}
\usepackage{bm}
\usepackage{color}

\begin{document}
\begin{center}
\begin{large}
{\bf Parameters of noncommutativity in Lie-algebraic noncommutative space}
\end{large}
\end{center}

\centerline {Kh. P. Gnatenko \footnote{E-Mail address: khrystyna.gnatenko@gmail.com}}
\medskip
\centerline {\small \it Ivan Franko National University of Lviv, Department for Theoretical Physics,}
\centerline {\small \it 12 Drahomanov St., Lviv, 79005, Ukraine}
\centerline {\small \it  Laboratory for Statistical Physics of Complex Systems}
\centerline {\small \it  Institute for Condensed Matter Physics, NAS of Ukraine, Lviv, 79011, Ukraine}

\abstract{We find condition on the parameters of noncommutativity on which a list of important results can be obtained in a space with Lie-algebraic noncommutativity. Namely, we show that the weak equivalence principle is recovered in the space, the Poisson brackets for coordinates and momenta of the center-of-mass of a composite system do not depend on its composition and reproduce relations of noncommutative algebra for coordinates and momenta of  individual particles  if parameters of noncommutativity corresponding to a particle are proportional inversely to its mass. In addition in particular case of Lie-algebraic noncommutativity (space coordinates commute to time) on this condition the motion of the center-of-mass is independent of the relative motion and problem of motion of the center-of-mass and problem corresponding to the internal motion can be studied separately.

Key words: noncommutative space, weak equivalence principle, composite system
PACS numbers: 11.10 Nx, 11.90.+t
}

\section{Introduction}

Idea that space coordinates may not commute was suggested by Heisenberg and formalized by Snyder \cite{Snyder}. Growing of interest to studies of noncommutative space is connected to development of String Theory and Quantum Gravity (see, for instance, \cite{Witten,Doplicher}).

Noncommutative algebra of Lie-type is characterized by the following commutation relations for coordinates
  \begin{eqnarray}
 [X_{i},X_{j}]=i\hbar\theta^{k}_{ij}X_k,\label{0}
 \end{eqnarray}
where $\theta^k_{ij}$ are constants called parameters of noncommutativity which are antisymmetric to lower indexes and can be particularly chosen (see, for instance, \cite{DaszkiewiczPRD,Lukierski,Lukierski18}).
In the present paper we consider different types of algebras with Lie-algebraic noncommutativity of space coordinates which correspond to particular choices of parameters $\theta^k_{ij}$ in (\ref{0}). Namely, noncommutative algebras with space coordinates commuting to time, space coordinates commuting to space \cite{DaszkiewiczPRD} and their generalizations \cite{Miao} are considered. We propose condition on the parameters of noncommutativity  which gives possibility to solve a list of problems in a space with Lie-algebraic noncommutativity, among them problem of violation of the weak equivalence principle and problem of description of motion of the center-of-mass of a many-particle system.

Studies of problems of many particles in the frame of noncommutative algebras give possibility to examine effect of space quantization on the properties of a wide class of physical systems including macroscopic ones.
Systems of many particles were examined in a space with canonical noncommutativity of coordinates ($[X_i,X_j]=i\hbar\theta_{ij}$, $\theta_{ij}$ are constants) \cite{Ho,Bellucci,GnatenkoPLA13}, in a phase space with canonical noncommutativity of coordinates and momenta ($[X_i,X_j]=i\hbar\theta_{ij}$, $[P_i,P_j]=i\hbar\eta_{ij}$, $\theta_{ij}$, $\eta_{ij}$ are constants)\cite{Djemai,GnatenkoPLA17}, noncommutative phase space of canonical type with rotational symmetry \cite{GnatenkoIJMPA18}.
In canonically deformed space-time a set of interacting harmonic oscillators and a system of particles moving in the gravitational field were studied in \cite{Daszkiewicz}. In \cite{Daszkiewicz18} the author examined energy levels of a system of two particles in Coulomb potential  in the case when  spatial coordinates commute to time-dependent function (twist-deformed space-time). In twisted N-enlarged Newton-Hooke space-time the quantum model of many particles was studied in \cite{Daszkiewicz1}. As an example a system of particles moving in Coulomb potential was examined.

In the present paper we study a problem of description of motion of the center-of-mass of composite system made of $N$ particles in Lie-algebraic noncommutative space in general case when coordinates of different particles satisfy noncommutative algebra of Lie-type with different parameters of noncommutativity. It is shown that the Poisson brackets for coordinates and momenta of the center-of-mass do not reproduce relations of noncommutative algebra for coordinates and momenta of individual particles. The motion of the center-of-mass of composite system depends on the relative motion.

 Implementation of the equivalence principle was examined in a space with coordinates noncommutativity of canonical type \cite{GnatenkoPLA13,Saha,Saha1}, in noncommutative phase space of canonical type \cite{Bastos1,Bertolami2,GnatenkoPLA17}, in rotationally-invariant noncommutative phase space \cite{GnatenkoEPL18}.

In the present paper  we study the weak equivalence principle in a space with Lie-algebraic noncommutativity. For this purpose a motion of a particle (macroscopic body) in gravitational field is examined.  We show that if only one condition on the parameters of noncommutativity is satisfied (if parameters of noncommutativity corresponding to a particle are proportional inversely to its mass) the weak equivalence principle is recovered in the noncommutative space, noncommutative algebra for coordinates and momenta  of the center-of-mass reproduce noncommutative algebra for coordinates and momenta of individual particles.

The paper is organized as follows. In Section 2 we study many-particle system in a space with Lie-algebraic noncommutativity and propose condition on the parameters of noncommutativity on which commutation relations for coordinates of the center-of-mass correspond to noncommutative algebra of Lie type. In Section 3 we show that the same condition on the parameters of noncommutativity is important for recovering of the weak equivalence principle in the space. Conclusions are presented in Section 4.

\section{Many-particle system in Lie-algebraic noncommutative space and parameters of noncommutativity}

 In general case coordinates and momenta which correspond to different particles may satisfy noncommutative algebra with different parameters. So, there is a problem of description of motion of the center-of-mass of composite system in the noncommutative space. It is important to find parameters of noncommutativity which describe this motion.

Let us first examine particular case of Lie-algebraic noncommutativity, namely the case of space coordinates commuting to time  \cite{DaszkiewiczPRD,DaszkiewiczMPLA08}.
The corresponding noncommutative algebra reads
 \begin{eqnarray}
 [X_i,X_j]=\frac{i\hbar t}{\kappa}\left(\delta_{i \rho}\delta_{j\tau}-\delta_{i\tau}\delta_{j\rho}\right),{}\label{nt}\\{}
  [X_i,P_j]=i\hbar\delta_{ij},{}\\{}
  [P_i,P_j]=0,\label{nt1}
 \end{eqnarray}
with $\kappa$ being  mass-like parameter, indexes $i,j=(1,2,3)$.  Indexes $\rho$, $\tau$ are fixed and different. For instance if one chooses $\rho=1$, $\tau=2$ one has $[X_1,X_2]=i\hbar t/\kappa$, $[X_2,X_3]=[X_1,X_3]=0$.
In the classical limit $\hbar\rightarrow0$ from (\ref{nt})-(\ref{nt1}) one obtains the following Poisson brackets \cite{DaszkiewiczPRD}
 \begin{eqnarray}
 \{X_i,X_j\}=\frac{t}{\kappa}\left(\delta_{i \rho}\delta_{j\tau}-\delta_{i\tau}\delta_{j\rho}\right),{}\label{tn}\\{}
  \{X_i,P_j\}=\delta_{ij},{}\\{}
  \{P_i,P_j\}=0.\label{tn1}
 \end{eqnarray}

In general case coordinates of different particles may satisfy noncommutative algebra with different parameters $\kappa$. So, we can write
\begin{eqnarray}
\{X_{i}^{(a)},X_{j}^{(b)}\}=\frac{ t}{\kappa_a}\left(\delta_{i \rho}\delta_{j\tau}-\delta_{i\tau}\delta_{j\rho}\right)\delta_{ab},\label{al0}\\{}
\{X_{i}^{(a)},P_{j}^{(b)}\}=\delta_{ab}\delta_{ij},\\{}
\{P_{i}^{(a)},P_{j}^{(b)}\}=0,\label{al1}
\end{eqnarray}
where indexes $a$ and $b$ label the particles.
Let us consider a system of $N$ particles of masses $m_a$ and introduce the coordinates  and momenta of the center-of-mass and coordinates and momenta of the relative motion
 \begin{eqnarray}
 \tilde{{\bf P}}=\sum_{a}{\bf P}^{(a)}, \ \ \tilde{{\bf X}}=\sum_{a}\mu_{a}{\bf X}^{(a)},\label{05}\\
\Delta{\bf P}^{{a}}={\bf P}^{(a)}-\mu_{a}\tilde{{\bf P}},\ \ {\Delta\bf X}^{(a)}={\bf X}^{(a)}-\tilde{{\bf X}},\label{06}
\end{eqnarray}
with $\mu_a=m_{a}/M$, $M=\sum_{a}m_a$ and $X^{(a)}_i$, $P^{(a)}_i$ satisfying (\ref{al0})-(\ref{al1}). Taking into account (\ref{al0})-(\ref{al1}) and (\ref{05})-(\ref{06}), we obtain
\begin{eqnarray}
\{\tilde{X}_i,\tilde{X}_j\}= t\sum_{a}\frac{\mu_{a}^{2}}{\kappa_a}\left(\delta_{i \rho}\delta_{j\tau}-\delta_{i\tau}\delta_{j\rho}\right),\label{07}\\{}
\{\tilde{X}_i,\tilde{P}_j\}=\delta_{ij}, \ \ \{\tilde{P}_i,\tilde{P}_j\}=0 \label{08}\label{007}\\{}
\{\Delta{X}_i^{(a)},\Delta{X}_j^{(b)}\}= t\left(\frac{\delta^{ab}}{\kappa_a}-\frac{\mu_{a}}{\kappa_{a}}-\frac{\mu_{b}}{\kappa_{b}}+\sum_{c}\frac{\mu_{c}^{2}}{\kappa_c}\right)\left(\delta_{i \rho}\delta_{j\tau}-\delta_{i\tau}\delta_{j\rho}\right),\\{}
\{\Delta{X}^{(a)}_i,\Delta{P}^{(b)}_j\}=\delta_{ab}-\mu_b,\\{}
\{\Delta{X}_i^{(a)},\tilde{X}_j\}= t \left(\frac{\mu_{a}}{\kappa_{a}}-\sum_{c}\frac{\mu_{c}^{2}}{\kappa_c}\right)\left(\delta_{i \rho}\delta_{j\tau}-\delta_{i\tau}\delta_{j\rho}\right),\label{cr}\\{}
\{\Delta{P}_i^{(a)},\Delta{P}_i^{(b)}\}=\{\tilde{P}_i,\Delta{P}_j^{(b)}\}=0.{}\label{09}
\end{eqnarray}
From (\ref{07}), (\ref{007}) we have that coordinates of the center-of-mass satisfy noncommutative algebra with effective parameter of noncommutativity
\begin{eqnarray}
{\tilde{\theta}}^{0}_{ij}=\sum_{a}\frac{\mu_{a}^{2}}{\kappa_a}\left(\delta_{i \rho}\delta_{j\tau}-\delta_{j\tau}\delta_{i\rho}\right)=\frac{1}{{\kappa}_{eff}}\left(\delta_{i \rho}\delta_{j\tau}-\delta_{i\tau}\delta_{j\rho}\right)\label{eff},
\end{eqnarray}
which depends on masses of particles which form the system and parameters $\kappa_a$. So, it depends on the composition of the system. Here we introduce notation
\begin{eqnarray}
\frac{1}{{\kappa}_{eff}}=\sum_{a}\frac{\mu_{a}^{2}}{\kappa_a}\label{tk}.
\end{eqnarray}
 In particular case of a system of particles with masses $m_1=m_2=..=m_N=m$ and parameters $\kappa_1=\kappa_2=...=\kappa_N=\kappa$ one has
\begin{eqnarray}
\tilde{\theta}^0_{ij}=\frac{1}{N\kappa}\left(\delta_{ik}\delta_{jl}-\delta_{jk}\delta_{il}\right)\label{eff1}.
\end{eqnarray}
So, there is reduction of effective parameter of noncommutativity with respect to parameters of noncommutativity corresponding to individual particles. This parameter decreases with increasing of number of particles in the system as $1/N$ (\ref{eff1}).

It is worth to stress that according to (\ref{cr}) the Poisson brackets for coordinates of the center-of-mass and coordinates of the relative motion do not vanish. So, the motion of the center-of-mass is not independent of the relative motion in a space (\ref{nt})-(\ref{nt1}).

Note that in the case when
\begin{eqnarray}
\frac{\kappa_a}{m_a}=\gamma_{\kappa}=const,\label{cond}
\end{eqnarray}
where $\gamma_{\kappa}$ is a constant which is the same for particles with different masses, we have that the Poisson brackets for coordinates of the center-of-mass  and coordinates of the relative motion are equal to zero $\{\Delta{X}_i^{(a)},\tilde{X}_j\}=0$.  As a result one can consider problems for motion of the center-of-mass of a system and problem for internal motion in the system independently. Also if condition (\ref{cond}) holds effective parameter of noncommutativity does not depend on composition of the system
\begin{eqnarray}
\tilde{\theta}^0_{ij}=\frac{1}{\gamma_{\kappa}M}\left(\delta_{i \rho}\delta_{j\tau}-\delta_{j\tau}\delta_{i\rho}\right)\label{eff2}.
\end{eqnarray}
So, the Poisson brackets for coordinates of the center-of-mass of a composite system do not depend on its composition and are determined by its total mass. Note that comparing (\ref{eff2}) with (\ref{eff}) we have ${\kappa_{eff}}/M=\gamma_{\kappa}$. So, condition  (\ref{cond}) is satisfied also for ${\kappa}_{eff}$. Effective parameter of noncommutativity (\ref{eff2}) is proportional inversely to the total mass of the system.

Let us consider the case when space coordinates commute to space
 \begin{eqnarray}
 \{X^{(a)}_k,X^{(b)}_{\gamma}\}=\delta_{ab}\frac{X^{(a)}_l}{\tilde{\kappa}},{}\ \  \{X^{(a)}_l,X^{(b)}_{\gamma}\}=-\delta_{ab}\frac{X^{(a)}_k}{\tilde{\kappa}}, \ \ \{X^{(a)}_k,X^{(b)}_{l}\}=0, \label{sts}\\{}
 \{P^{(a)}_k,X^{(b)}_{\gamma}\}=\delta_{ab}\frac{P^{(a)}_l}{\tilde{\kappa}},{}\ \ \{P^{(a)}_l,X^{(b)}_{\gamma}\}=-\delta_{ab}\frac{P^{(a)}_k}{\tilde{\kappa}}, \label{sts3} \\ \{X^{(a)}_i,P^{(b)}_j\}=\delta_{ab}\delta_{ij}, \ \ \{X^{(a)}_{\gamma},P^{(b)}_{\gamma}\}=\delta_{ab},{}\ \
 \{P^{(a)}_m,P^{(b)}_n\}=0,{} \label{sts1}
 \end{eqnarray}
where indexes $k$, $l$, $\gamma$ are different and fixed, $k,l,\gamma=(1,2,3)$,  $i\neq\gamma$, $j\neq\gamma$ and $m,n=(1,2,3)$, $\tilde{\kappa}$ is a constant \cite{DaszkiewiczPRD}. Indexes $a,b$ label the particles.
Introducing coordinates and momenta of the center-of-mass and coordinates and momenta of the relative motion 
(\ref{05})-(\ref{06}) with $X^{(a)}_i$, $P^{(a)}_i$ satisfying (\ref{sts})-(\ref{sts1})
one can find
 \begin{eqnarray}
\{\tilde{X}_k,\tilde{X}_{\gamma}\}=\sum_{a}\frac{\mu_{a}^{2}X^{(a)}_l}{\tilde{\kappa}_a}, \  \ \{\tilde{X}_l,\tilde{X}_{\gamma}\}=-\sum_{a}\frac{\mu_{a}^{2}X^{(a)}_k}{\tilde{\kappa}_a},\ \ \{\tilde{X}_k,\tilde{X}_l\}=0.\label{cm}\\ {}
\{\tilde{P}_k,\tilde{X}_{\gamma}\}=\sum_{a}\frac{\mu_{a}P^{(a)}_l}{\tilde{\kappa}_a},{}\ \  \{\tilde{P}_l,\tilde{X}_{\gamma}\}=-\sum_{a}\frac{\mu_{a}P^{(a)}_k}{\tilde{\kappa}_a}, \label{xp1}\\{}   \{\tilde{X}_i,\tilde{P}_j\}=\delta_{ij}, \ \ \{\tilde{X}_{\gamma},\tilde{P}_{\gamma}\}=1{}\ \
\{\tilde{P}_m,\tilde{P}_n\}=0.{} \label{cm1}
\end{eqnarray}
Note that the relations for coordinates of the center-of-mass (\ref{cm}) do not reproduce relations of noncommutative algebra (\ref{sts}). The relations are determined by the expressions $\sum_{a}{\mu_{a}^{2}X^{(a)}_l}/{\tilde{\kappa}_a}$, $\sum_{a}{\mu_{a}^{2}X^{(a)}_k}/{\tilde{\kappa}_a}$ which depend on the composition of the system and on the coordinates of particles in the system. The same conclusion can be done for  relations (\ref{xp1}). These relations do not reproduce relations (\ref{sts3}) of noncommutative algebra (\ref{sts})-(\ref{sts1}). It is worth stressing that if condition
\begin{eqnarray}
\frac{\tilde{\kappa}_a}{m_a}=\gamma_{\tilde{\kappa}}=const,\label{cond1}
\end{eqnarray}
 is satisfied one has
 \begin{eqnarray}
\{\tilde{X}_k,\tilde{X}_{\gamma}\}=\frac{1}{\tilde{\kappa}_{eff}}\tilde{X}_{l}, \ \ \{\tilde{X}_l,\tilde{X}_{\gamma}\}=-\frac{1}{\tilde{\kappa}_{eff}}\tilde{X}_{k},\ \ \{\tilde{X}_k,\tilde{X}_l\}=0,\label{cmc}\\{}
\{\tilde{P}_k,\tilde{X}_{\gamma}\}= \frac{\tilde{P}_l}{\tilde{\kappa}_{eff}},{}\ \  \{\tilde{P}_l,\tilde{X}_{\gamma}\}=-\frac{\tilde{P}_k}{\tilde{\kappa}_{eff}},\label{cmc1}
\end{eqnarray}
with $\tilde{\kappa}_{eff}=\gamma_{\tilde{\kappa}}M$. Here $\gamma_{\tilde{\kappa}}$ is a constant which does not depend on  mass. So, if condition (\ref{cond1}) holds in the right-hand side of relations (\ref{cmc}), (\ref{cmc1}) one has coordinates of the center-of-mass  and the total momenta, respectively. So, relations (\ref{cmc}), (\ref{cmc1})  reproduce relations of noncommutative algebra (\ref{sts}), (\ref{sts3}).

Note that in the space with noncommutativity (\ref{sts})-(\ref{sts1}) the motion of the center-of-mass is not independent on the relative motion
\begin{eqnarray}
\{\Delta{X}_k^{(a)},\tilde{X}_{\gamma}\}=\{\tilde{X}_k,\Delta{X}^{(a)}_{\gamma}\}= \frac{\mu_{a}X^{(a)}_l}{\tilde{\kappa}_{a}}-\sum_{b}\frac{\mu_{b}^{2}X^{(b)}_l}{\tilde{\kappa}_b},\\{} \{\Delta{X}_l^{(a)},\tilde{X}_{\gamma}\}=\{\tilde{X}_l,\Delta{X}^{(a)}_{\gamma}\}=- \frac{\mu_{a}X^{(a)}_k}{\tilde{\kappa}_{a}}+\sum_{b}\frac{\mu_{b}^{2}X^{(b)}_k}{\tilde{\kappa}_b}, \\{}
\{\Delta{X}_k^{(a)},\tilde{X}_l\}=\{\Delta{X}_l^{(a)},\tilde{X}_k\}=0,\\
\{\tilde{P}_k,\Delta{X}^{(a)}_{\gamma}\}=\frac{P^{(a)}_l}{\tilde{\kappa}_a}-\sum_{b}\frac{\mu_{b}P^{(b)}_l}{\tilde{\kappa}_b}, \ \  \{\tilde{P}_l,\Delta{X}^{(a)}_{\gamma}\}=-\frac{P^{(a)}_k}{\tilde{\kappa}_a}+\sum_{b}\frac{\mu_{b}P^{(b)}_k}{\tilde{\kappa}_b}, \label{xp3} \\ {}
\{\Delta{P}^{(a)}_k,\tilde{X}_{\gamma}\}=\mu_a\left(\frac{P^{(a)}_{l}}{\tilde{\kappa}_a}-\sum_{b}\frac{\mu_{b}P^{(b)}_l}{\tilde{\kappa}_b}\right), \ \  \{\Delta{P}^{(a)}_l,\tilde{X}_{\gamma}\}=-\mu_a\left(\frac{P^{(a)}_k}{\tilde{\kappa}_a}-\sum_{b}\frac{\mu_{b}P^{(b)}_k}{\tilde{\kappa}_b}\right), \\ {}
\{\tilde{P}_k,\Delta{X}^{(a)}_l\}=\{\tilde{P}_l,\Delta{X}^{(a)}_k\}=\{\Delta{P}^{(a)}_l,\tilde{X}_k\}=\{\Delta{P}^{(a)}_k,\tilde{X}_l\}=0, \label{xp4}
\end{eqnarray}
Even if condition (\ref{cond1}) is satisfied one has that these  relations can be rewritten in more simple form \begin{eqnarray}
\{\Delta{X}_k^{(a)},\tilde{X}_{\gamma}\}=\{\tilde{X}_k,\Delta{X}^{(a)}_{\gamma}\}=\frac{1}{\tilde{\kappa}_{eff}}\Delta{X}^{(a)}_l,\\{} \{\Delta{X}_{l}^{(a)},\tilde{X}_{\gamma}\}=\{\tilde{X}_l,\Delta{X}^{(a)}_{\gamma}\}=-\frac{1}{\tilde{\kappa}_{eff}}\Delta{X}^{(a)}_k, \\{}
\{\tilde{P}_{k},\Delta{X}^{(a)}_{\gamma}\}=\frac{1}{\tilde{\kappa}_a}\Delta{P}^{(a)}_l, \ \  \{\tilde{P}_l,\Delta{X}^{(a)}_{\gamma}\}=-\frac{1}{\tilde{\kappa}_a}\Delta{P}^{(a)}_k, \label{xp} \\ {}
\{\Delta{P}^{(a)}_k,\tilde{X}_{\gamma}\}=\frac{1}{\tilde{\kappa}_{eff}}\Delta{P}^{(a)}_l, \ \  \{\Delta{P}^{(a)}_l,\tilde{X}_{\gamma}\}=-\frac{1}{\tilde{\kappa}_{eff}}\Delta{P}^{(a)}_k,
\end{eqnarray}
but they do not vanish.

Similar conclusion can be done for the generalized algebras with Lie-type of noncommutativity determined  in \cite{Miao}. In more general case one can write
\begin{eqnarray}
 \{X_i,X_j\}=\theta^0_{ij}t+\theta^k_{ij}X_k,{}\label{gen}\\{}
  \{X_i,P_j\}=\delta_{ij}+\bar{\theta}^k_{ij}X_k+\tilde{\theta}^k_{ij}P_k,{}\\{}
  [P_i,P_j]=0,\label{gen1}
 \end{eqnarray}
 where $\theta^0_{ij}$, $\theta^k_{ij}$, $\bar{\theta}^k_{ij}$, $\tilde{\theta}^k_{ij}$ are antisymmetric to lower indexes parameters of noncommutativity, indexes $i,j,k=(1,2,3)$, the Poisson brackets for time and spatial coordinates vanish  \cite{Miao}. Due to the Jacobi identity there are constrains on the parameters   $\theta^0_{ij}$, $\theta^k_{ij}$, $\bar{\theta}^k_{ij}$, $\tilde{\theta}^k_{ij}$.  On the basis of these constrains the authors of paper \cite{Miao} determined two types of noncommutative algebras
\begin{eqnarray}
\{X_k,X_{\gamma}\}=-\frac{t}{\kappa}+\frac{X_l}{\tilde{\kappa}},{}\ \  \{X_l,X_{\gamma}\}=\frac{t}{\kappa}-\frac{X_k}{\tilde{\kappa}}, \ \ \{X_k,X_{l}\}=\frac{t}{\kappa}, \label{gsts}\\{}
\{P_k,X_{\gamma}\}=\frac{P_l}{\tilde{\kappa}},{}\ \ \{P_l,X_{\gamma}\}=-\frac{P_k}{\tilde{\kappa}},  \ \ \{X_i,P_j\}=\delta_{ij}, \ \   \{X_{\gamma},P_{\gamma}\}=1 {} \\
\{P_m,P_n\}=0,{} \label{gsts1}
\end{eqnarray}
which corresponds  to the case $\theta^0_{kl}=-\theta^0_{k\gamma}=1/\kappa$,  $\theta^0_{l\gamma}=1/\kappa$,  $\theta^l_{k\gamma}=-\theta^k_{l\gamma}=\tilde{\theta}^l_{k\gamma}=-\tilde{\theta}^k_{l\gamma}=1/\tilde{\kappa}$ and
\begin{eqnarray}
\{X_k,X_{\gamma}\}=-\frac{t}{\kappa}+\frac{X_l}{\tilde{\kappa}},{}\ \  \{X_l,X_{\gamma}\}=\frac{t}{\kappa}-\frac{X_k}{\tilde{\kappa}}, \ \ \{X_k,X_{l}\}=0, \label{gsts2}\\{}
\{P_k,X_{\gamma}\}=\frac{X_l}{\bar{\kappa}}+\frac{P_l}{\tilde{\kappa}},{}\ \ \{P_l,X_{\gamma}\}=\frac{X_k}{\bar{\kappa}}-\frac{P_k}{\tilde{\kappa}},  \ \ \{X_i,P_j\}=\delta_{ij}, \ \   \{X_{\gamma},P_{\gamma}\}=1,{}\\
\{P_m,P_n\}=0,{} \label{gsts12}
\end{eqnarray}
which corresponds  to  $\theta^0_{l\gamma}=-\theta^0_{k\gamma}=1/\kappa$,  $\theta^l_{k\gamma}=-\theta^k_{l\gamma}=1/\tilde{\kappa}$,  $\tilde{\theta}^l_{k\gamma}=-\tilde{\theta}^k_{l\gamma}=1/\tilde{\kappa}$, $\bar{\theta}^l_{k\gamma}=-\bar{\theta}^k_{l\gamma}=1/\bar{\kappa}$.
 These algebras  are generalizations of (\ref{al0})-(\ref{al1}), (\ref{sts})-(\ref{sts1}).

Coordinates and momenta of different particles may satisfy noncommutative algebra with different parameters. So, in general case we can write
\begin{eqnarray}
\{X^{(a)}_i,X^{(b)}_j\}=\delta_{ab}\theta^{0(a)}_{ij}t+\delta_{ab}\theta^{k(a)}_{ij}X^{(a)}_k,{} \label{c01}\\{}
\{X^{(a)}_i,P^{(b)}_j\}=\delta_{ab}\delta_{ij}+\delta_{ab}\bar{\theta}^{k(a)}_{ij}X^{(a)}_k+\delta_{ab}\tilde{\theta}^{k(a)}_{ij}P^{a}_k,{}\\{}
\{P^{(a)}_i,P^{(b)}_j\}=0.\label{c001}
 \end{eqnarray}
 Taking into account (\ref{c01})-(\ref{c001}) and (\ref{05})  we find
\begin{eqnarray}
\{\tilde{X}_i,\tilde{X}_j\}=\sum_a\mu_a^2\theta^{0(a)}_{ij}t+\sum_a\mu_a^2\theta^{k(a)}_{ij}X^{(a)}_k,{} \label{c11}\\{}
\{\tilde{X}_i,\tilde{P}_j\}=\delta_{ij}+\sum_{a}\mu_a\bar{\theta}^{k(a)}_{ij}X^{(a)}_k+\sum_a\mu_a\tilde{\theta}^{k(a)}_{ij}P^{a}_k,{}\\{}
\{\tilde{P}_i,\tilde{P}_j\}=0.\label{c1}
 \end{eqnarray}
 Note that the Poisson brackets for the coordinates and momenta of the center-of-mass depend on the parameters of noncommutativity which correspond to particles forming the system and on their masses.

 On the basis of results of previous studies in spaces (\ref{al0})-(\ref{al1}), (\ref{sts})-(\ref{sts1}) one can conclude, that if parameters of noncommutativity depend on mass as
\begin{eqnarray}
\theta^{0(a)}_{ij}m_a={\gamma^{0}_{ij}}=const,\ \ \theta^{k(a)}_{ij}m_a=\gamma^{k}_{ij}=const ,\ \
\tilde{\theta}^{k(a)}_{ij}m_a={\tilde{\gamma}^{k}_{ij}}=const,\label{cond2}
\end{eqnarray}
(here constants $\gamma^{0}_{ij}$, $\gamma^{k}_{ij}$, $\tilde{\gamma}^{k}_{ij}$  are the same for particles with different masses and are antisymmetric to lower indexes) and parameters $\bar{\theta}^{k(a)}_{ij}$ are the same for different particles
\begin{eqnarray}
\bar{\theta}^{k(a)}_{ij}=\bar{\theta}^{k}_{ij},\label{cond3}
\end{eqnarray}
the algebra for coordinates and momenta of the center-of-mass reproduce noncommutative algebra for the individual particles. One has
\begin{eqnarray}
\{\tilde{X}_i,\tilde{X}_j\}=\theta^{0(eff)}_{ij}t+\theta^{k(eff)}_{ij}\tilde{X}_k,{} \label{cm111}\\{}
\{\tilde{X}_i,\tilde{P}_j\}=\delta_{ij}+\bar{\theta}^{k}_{ij}\tilde{X}_k+\tilde{\theta}^{k(eff)}_{ij}\tilde{P}_k,{}\label{cm12}
 \end{eqnarray}
where $\theta^{0(eff)}_{ij}=\gamma^{0}_{ij}/M$, $\theta^{k(eff)}_{ij}=\gamma^{k}_{ij}/M$, $\tilde\theta^{k(eff)}_{ij}=\tilde{\gamma}^{k}_{ij}/M$ and $M=\sum_am_a$.

In particular cases (\ref{gsts})-(\ref{gsts1}), (\ref{gsts2})-(\ref{gsts12}) of noncommutative algebra (\ref{gen})-(\ref{gen1})  conditions (\ref{cond2}) can be rewritten as (\ref{cond}), (\ref{cond1}) and from condition (\ref{cond3}) follows
\begin{eqnarray}
\bar{\kappa}_a=\bar{\kappa}.\label{cond5}
\end{eqnarray}

Note that the results presented in this section can be easily generalized to the quantum case, considering the corresponding operators and commutation relations for them.

In the next section we show that on the same conditions on the parameters of noncommutativity (\ref{cond2}) the weak equivalence principle is recovered in a Lie-algebraic noncommutative space.

\section{Weak equivalence principle in Lie-algebraic noncommutative space and parameters of noncommutativity}

Let us examine implementation of the weak equivalence principle in Lie-algebraic noncommutative space.
For this purpose  we consider the motion of a particle of mass $m$ in gravitational field in the space and study the following Hamiltonian
 \begin{eqnarray}
H=\frac{{\bf P}^2}{2m}+mV(X_1,X_2,X_3),\label{h}
\end{eqnarray}
here $V=V(X_1,X_2,X_3)$ describes the field. Note that in Hamiltonian (\ref{h}) the inertial mass in the first term is equal to the gravitational mass  in the second term.

We would like to note here that classical mechanics in a space with noncommutativity of Lie type was studied in \cite{DaszkiewiczPRD,Miao}. The authors examined Newton equations for a particle in a constant external field force in a spaces with noncommutative algebras (\ref{tn})-(\ref{tn1}), (\ref{sts})-(\ref{sts1}) \cite{DaszkiewiczPRD} and (\ref{gsts})-(\ref{gsts1}), (\ref{gsts2})-(\ref{gsts12}) \cite{Miao}.

Let us first consider a space with Lie-algebraic structure in the case when space coordinates commuting to time  (\ref{tn})-(\ref{tn1}).
 So, taking into account (\ref{h}) and (\ref{tn})-(\ref{tn1}), we can write
 \begin{eqnarray}
\dot{X}_i=\{X_i,H\}=\frac{P_i}{m}+\frac{tm}{\kappa}\frac{\partial V}{\partial X_k}\left(\delta_{i\rho}\delta_{k\tau}-\delta_{i\tau}\delta_{k\rho}\right),\label{xcd}\\
\dot{P}_i=\{P_i,H\}=-m\frac{\partial V}{\partial X_i},
 \end{eqnarray}
 see also \cite{DaszkiewiczPRD}, where Hamilton equations were presented for arbitrary field. Note that because of nooncommutativity one has additional term in (\ref{xcd}) which is proportional to the mass $m$.
According to the weak equivalence principle the kinematic characteristics (velocity and position) of a point
mass in a gravitational field depend only on its initial position and velocity, and are
independent of its mass, composition and structure. So, from (\ref{xcd}) one can state that the weak equivalence principle is violated because of noncommutativity of coordinates (\ref{tn}).

It is important to note that if condition (\ref{cond}) is satisfied one has
 \begin{eqnarray}
\dot{X}_i=P^{\prime}_i+\frac{t}{\gamma_{\kappa}}\frac{\partial V}{\partial X_k}\left(\delta_{i\rho}\delta_{k\tau}-\delta_{i\tau}\delta_{k\rho}\right),\label{xdc}\\
\dot{P}^{\prime}_i=-\frac{\partial V}{\partial X_i},\label{xdc1}
 \end{eqnarray}
here we use notation $P^{\prime}_i=P_i/m$. Note that equations  (\ref{xdc}), (\ref{xdc1}) do not contain mass and depend on the constant $\gamma_{\kappa}$ which is the same for particles with different masses. So,  we can conclude that  their solutions $X_i(t)$, $P^{\prime}_i(t)$ do not depend on the mass. Therefore the weak equivalence principle is recovered due to condition (\ref{cond}).

Let us study implementation of the weak equivalence principle in the space described by relations (\ref{sts})-(\ref{sts1})(space coordinates commute to space). Taking into account (\ref{h}) and  (\ref{sts})-(\ref{sts1}) one has
\begin{eqnarray}
\dot{X}_k=\frac{P_k}{m}+\frac{m X_{l}}{\tilde{\kappa}}\frac{\partial V}{\partial X_{\gamma}},\ \ \dot{X}_{l}=\frac{P_l}{m}-\frac{m X_{k}}{\tilde{\kappa}}\frac{\partial V}{\partial X_{\gamma}},\label{xxd}\\
\dot{X}_{\gamma}=\frac{P_{\gamma}}{m}-\frac{m X_{l}}{\tilde{\kappa}}\frac{\partial V}{\partial X_{k}}+\frac{m X_{k}}{\tilde{\kappa}}\frac{\partial V}{\partial X_{l}},\\
\dot{P}_k=-m\frac{\partial V}{\partial X_k}+\frac{m P_{l}}{\tilde{\kappa}}\frac{\partial V}{\partial X_{\gamma}},\ \
\dot{P}_l=-m\frac{\partial V}{\partial X_l}-\frac{m P_{k}}{\tilde{\kappa}}\frac{\partial V}{\partial X_{\gamma}},\\
\dot{P}_{\gamma}=-m\frac{\partial V}{\partial X_{\gamma}},\label{xxd1}
\end{eqnarray}
see also \cite{DaszkiewiczPRD}. From (\ref{xxd})-(\ref{xxd1}) we have that if we assume that the parameters $\tilde{\kappa}$ are the same for particles with different masses the velocity of a particle in gravitational field  in a space (\ref{sts})-(\ref{sts1}) depends on its mass. So, relations (\ref{sts})-(\ref{sts1}) lead to violation of the weak equivalence principle.

We would like to mention that if condition (\ref{cond1}) is satisfied one has
\begin{eqnarray}
\dot{X}_k=P^{\prime}_k+\frac{X_{l}}{\gamma_{\tilde{\kappa}}}\frac{\partial V}{\partial X_{\gamma}},\ \
\dot{X}_{l}={P_l^{\prime}}-\frac{X_{k}}{\gamma_{\tilde{\kappa}}}\frac{\partial V}{\partial X_{\gamma}},\label{xxdc}\\
\dot{X}_{\gamma}={P^{\prime}_{\gamma}}-\frac{X_{l}}{\gamma_{\tilde{\kappa}}}\frac{\partial V}{\partial X_{k}}+\frac{ X_{k}}{\gamma_{\tilde{\kappa}}}\frac{\partial V}{\partial X_{l}},\\
\dot{P}^{\prime}_k=-\frac{\partial V}{\partial X_k}+\frac{P^{\prime}_{l}}{\gamma_{\tilde{\kappa}}}\frac{\partial V}{\partial X_{\gamma}},\ \
\dot{P}^{\prime}_l=-\frac{\partial V}{\partial X_l}-\frac{ P^{\prime}_{k}}{\gamma_{\tilde{\kappa}}}\frac{\partial V}{\partial X_{\gamma}},\\
\dot{P}^{\prime}_{\gamma}=-\frac{\partial V}{\partial X_{\gamma}}.\label{xxdc1}
\end{eqnarray}
Equations (\ref{xxdc})-(\ref{xxdc1}) do not depend on the mass, therefore their solutions  $X_k(t)$, $X_l(t)$, $X_{\gamma}(t)$, $P^{\prime}_k(t)$, $P^{\prime}_l(t)$, $P^{\prime}_{\gamma}(t)$ do not depend on the mass too. So, we can conclude that if condition on the parameter of noncommutativity (\ref{cond1}) holds the weak equivalence principle is preserved in a space with Lie-algebraic noncommutativity (\ref{sts})-(\ref{sts1}).

In more general case of noncommutative algebra of Lie-type (\ref{gen})-(\ref{gen1}) for a particle in gravitational field (\ref{h}) we have the following equations of motion
\begin{eqnarray}
\dot{X}_i=\frac{P_i}{m}+\bar{\theta}^k_{ij}\frac{P_j X_k}{m}+\tilde{\theta}^k_{ij}\frac{P_j P_k}{m}+m(\theta^{0}_{ij}t+\theta^{k}_{ij}X_k)\frac{\partial V}{\partial X_j},\label{xd}\\
\dot{P}_i=-m\frac{\partial V}{\partial X_i}-m(\bar{\theta}^{k}_{ij}X_{k}+\tilde{\theta}^{k}_{ij}P_k)\frac{\partial V}{\partial X_j}.
 \end{eqnarray}
In the case when parameters $\theta^{0}_{ij}$, $\theta^{k}_{ij}$, $\tilde{\theta}^{k}_{ij}$ are proportional inversely to mass (\ref{cond2}) and parameters $\bar{\theta}^{k}_{ij}$ are the same for particles with different masses (\ref{cond3}) we have
\begin{eqnarray}
\dot{X}_i={P^{\prime}_i}+\bar{\theta}^k_{ij}{P^{\prime}_j X_k}+\tilde{\gamma}^k_{ij}{P^{\prime}_j P^{\prime}_k}+(\gamma^{0}_{ij}t+\gamma^{k}_{ij}X_k)\frac{\partial V}{\partial X_j},\label{eq1}\\
\dot{P}^{\prime}_i=-\frac{\partial V}{\partial X_i}-(\bar{\theta}^{k}_{ij}X_{k}+\tilde{\gamma}^{k}_{ij}P^{\prime}_k)\frac{\partial V}{\partial X_j},\label{eq2}
 \end{eqnarray}
where $P^{\prime}_i=P_i/m$. From (\ref{eq1}), (\ref{eq2})  we can conclude that $X_i(t)$ and $P^{\prime}_i(t)$ do not depend on the mass of a particle. So, in a space with Lie-algebraic noncommutativity (\ref{gen})-(\ref{gen1}) the weak equivalence principle can be recovered due to the conditions (\ref{cond2}), (\ref{cond3}).

At the end of this section let us discuss more general case of motion of macroscopic body (composite system) of mass $M$ in gravitational field in noncommutative space with Lie-algebraic noncommutativity and study the weak equivalence principle. So, let us consider the following Hamiltonian
\begin{eqnarray}
H=\frac{{\bf \tilde{P}}^2}{2M}+MV(\tilde{X}_1,\tilde{X}_2,\tilde{X}_3)+H_{rel},\label{body}
\end{eqnarray}
where $\tilde{X}_i$, $\tilde{P}_i$ are coordinates and momenta of the center-of-mass of a body, $V(\tilde{X}_1,\tilde{X}_2,\tilde{X}_3)$ describes the filed. Hamiltonian $H_{rel}$  corresponds to the relative motion and depends on the relative coordinates and relative momenta.

 Let us first study the case when commutator of coordinates closes to time (\ref{nt})-(\ref{nt1}). As was shown in the previous section coordinates and momenta of the center-of-mass, coordinates and momenta of the relative motion  satisfy the following relations (\ref{07})-(\ref{09}). In the previous section we also concluded that if condition (\ref{cond}) is satisfied one has $\{\Delta X^{(a)}_i,\tilde{X}_j\}=0$ therefore
\begin{eqnarray}
\{\frac{{\bf \tilde{P}}^2}{2M}+MV(\tilde{X}_1,\tilde{X}_2,\tilde{X}_3),H_{rel}\}=0,
\end{eqnarray}
and we can consider the motion of the center-of-mass independently of the relative motion. Taking into account (\ref{body}) and  (\ref{07})-(\ref{007}), equations of motion for the center-of-mass of a body in gravitational field read
\begin{eqnarray}
\dot{{\tilde X}}_i=\frac{\tilde{P}_i}{M}+tM\sum_{a}\frac{\mu_{a}^{2}}{\kappa_a}\left(\delta_{i \rho}\delta_{j\tau}-\delta_{i\tau}\delta_{j\rho}\right)\frac{\partial V}{\partial \tilde{X}_j},\label{zxd3}\\
\dot{{\tilde P}}_i=-M\frac{\partial V}{\partial {\tilde X}_i}.
 \end{eqnarray}
Note that besides dependence of velocity on the mass of a body $M$, from  (\ref{zxd3}) one also has that the velocity of a body depends on the masses of a particles $m_a$ which form the body and on the parameters $\kappa_a$, so it depends on its composition. This also violates the weak equivalence principle. Taking into account (\ref{cond}) and using notation $\tilde{P}^{\prime}_i=\tilde{P}_i/M$, we obtain
\begin{eqnarray}
\dot{{\tilde X}}_i=\tilde{P}^{\prime}_i+t\sum_{a}\frac{1}{\gamma_\kappa}\left(\delta_{i \rho}\delta_{j\tau}-\delta_{i\tau}\delta_{j\rho}\right)\frac{\partial V}{\partial \tilde{X}_j},\label{zxd5}\\
\dot{{\tilde P}}^{\prime}_i=-\frac{\partial V}{\partial {\tilde X}_i}.\label{zxd6}
 \end{eqnarray}
From (\ref{zxd5}), (\ref{zxd6}) we can conclude that ${\tilde X}_i(t)$, $\tilde{P}^{\prime}_i(t)$ do not depend on the mass and composition of a body. So, the weak equivalence principle can be recovered due to condition (\ref{cond}).

In general case of noncommutative algebra (\ref{gen})-(\ref{gen1}), considering the case when influence of relative motion on  the motion of the center-of-mass can be neglected and taking into account conditions on the parameters of noncommutativity (\ref{cond2}), (\ref{cond3}), the equations for the center-of-mass of a body in gravitational field read
\begin{eqnarray}
\dot{{\tilde X}}_i={\tilde{P}^{\prime}_i}+\left(\bar{\theta}^{k}_{ij}\tilde{X}_k+\tilde{\gamma}^{k}_{ij}\tilde{P}^{\prime}_k\right) \tilde{P}^{\prime}_j+\left(\gamma^{0}_{ij}t+\gamma^{k}_{ij}\tilde{X}_k\right)\frac{\partial V}{\partial \tilde{X}_j},\label{eq111}\\
\dot{{\tilde P}}^{\prime}_i=-\frac{\partial V}{\partial {\tilde X}_i}-\left(\bar{\theta}^{k}_{ij}\tilde{X}_k+\tilde{\gamma}^{k}_{ij}\tilde{P}^{\prime}_k\right)\frac{\partial V}{\partial\tilde{X}_j},\label{eq211}
 \end{eqnarray}
 where $\tilde{P}^{\prime}_i=\tilde{P}_i/M$. Note, that the obtained equations depend on the constants $\gamma^{0}_{ij}$, $\gamma^{k}_{ij}$, $\tilde{\gamma}^{k}_{ij}$  and do not depend on mass and composition of the body. So, the weak equivalence principle is recovered in a space with Lie-algebraic noncommutativity if parameters of noncommutativity are determined by mass as (\ref{cond2}), (\ref{cond3}) and parameters $\bar{\theta}^{k}_{ij}$ are the same for particles of different masses and compositions.

\section{Conclusion}

Space with Lie-algebraic noncommutativity has been considered. We have studied a general case when different particles feel noncommutativity with different parameters and examined a problem of description of motion of composite system in different cases of Lie-algebraic noncommutativity  (space coordinates commute to time (\ref{nt})-(\ref{nt1}), space coordinates commute to space (\ref{sts})-(\ref{sts1}), noncommutative algebra of Lie type characterized by (\ref{gen})-(\ref{gen1})). We have shown that Poisson brackets for coordinates and momenta of the center-of-mass of a system depend on its composition and do not reproduce relations of noncommutative algebra for coordinates and momenta of individual particles. We have also concluded that because of noncommutativity the motion of the center-of-mass of composite system is effected by the relative motion.

Motion of a particle (a body) in gravitational field has been examined in a space with Lie-algebraic noncommutativity and the weak equivalence principle has been studied. We have concluded that noncommutativity causes additional terms in the equations of motion which depend on the mass of a particle (a body) in gravitational filed and on its composition. Therefore the weak equivalence principle is violated because of noncommutativity.

We have proposed condition on the parameters of noncommutativity which gives a possibility to obtain  important results in a space with Lie-algebraic structure. Namely, we have proposed to consider parameters of noncommutative algebra (\ref{gen})-(\ref{gen1})  $\theta^0_{ij}$, $\theta^k_{ij}$, $\tilde{\theta}^k_{ij}$  to be proportional inversely to  mass (\ref{cond2}) and parameters $\bar{\theta}^{k}_{ij}$ to be the same for different particles (\ref{cond3}). In this case Poisson brackets for coordinates and momenta of the center-of-mass of composite system do not depend on its composition and reproduce relations of noncommutative algebra for coordinates and momenta of individual particles. In particular cases of noncommutative algebra of Lie type (\ref{nt})-(\ref{nt1}), (\ref{sts})-(\ref{sts1}) condition (\ref{cond2}) reduces to (\ref{cond}), (\ref{cond1}), respectively. We have also shown that in the case of noncommutative algebra  (\ref{nt})-(\ref{nt1}) Poisson brackets for coordinates and momenta of the center-of-mass vanish due to condition (\ref{cond}). As a result the problem of motion of the center-of-mass and problem of the relative motion can be considered as independent problems.

Besides we have concluded that on the same conditions (\ref{cond2}), (\ref{cond3}) the motion of a body in gravitational field does not depend on its mass and composition and the weak equivalence principle is preserved in a space with Lie-algebraic noncommutativity.

So, due to assumption that parameters  $\theta^0_{ij}$, $\theta^k_{ij}$, $\tilde{\theta}^k_{ij}$ of noncommutative algebra are proportional inversely to the mass one can solve a list of problems in noncommutative space with Lie-algebraic noncommutativity. The importance of relations (\ref{cond2}) can be justified by the number of results which can be obtained due to them.

It is worth noting that relation of parameters of noncommutativity (parameters of deformation) with mass is important for description of composite system motion and for preserving of the weak equivalence principle in a space with canonical noncommutativity of coordinates \cite{GnatenkoPLA13,GnatenkoPLA17,GnatenkoMPLA17}, in rotationally-invariant noncommutative space \cite{GnatenkoIJMPA18,GnatenkoEPL18}, in deformed space with minimal length \cite{Tkachuk,Quesne,Tkachuk1}. So, idea to consider parameters of deformed algebra to be determined by mass is important for solving fundamental problems in a quantum space.

\section*{Acknowledgements}
The author thanks Prof. V. M. Tkachuk for his
advices and  support during research. 
This work was partly supported by the Project $\Phi\Phi$-63Hp
(No. 0117U007190) from the Ministry of Education
and Science of Ukraine and the grant of the President of Ukraine for support of scientific researches of young scientists (F-75).

\end{document}